**Demographic Patterns in Cybersecurity Culture: Insights from a Global Organisation Supporting Safety-Critical and Critical Infrastructure Sectors**

Tita Alissa Bach[1]* and Amandine Kaiser[1]

[1]Group Research and Development, DNV, Veritasveien 1, Høvik, Norway

*Corresponding author: Tita.Alissa.Bach@dnv.com

**Abstract (250 words):**

**Purpose**
This study investigates demographic differences in cybersecurity culture (CSC) in a large global organisation supporting safety-critical and critical-infrastructure sectors to target CSC improvement.

**Design/methodology/approach**

A global CSC survey was administered to all internal and external employees (N=21148), with 6,502 responses (30.75%). The questionnaire evaluates nine CSC dimensions such as Password Management, Governance, Email Use. Anonymous survey responses were analysed using Kruskal–Wallis tests and Dunn's post-hoc comparisons to identify differences across demographic variables (i.e., employment, recruitment paths, managerial role, gender, age, tenure, and work base).

**Findings**
CSC was broadly consistent across the organisation, with statistically significant but small to moderate demographic effects. CSC variations were observed across employment, age, recruitment paths, and line managerial role. In general, fulltime, internal, permanent, older employees (50+), M&A (Merge and Acquisition) recruits, and line managers consistently scored higher across multiple CSC dimensions. Parttime, younger, external employees, and those with 6–20 years of tenure in general scored lower. These patterns highlight higherscoring groups that may act as CSC carriers and lowerscoring groups that may benefit from tailored improvement measures, enabling organisational learning.

**Originality/value**
To the authors' knowledge, this study is among the first to provide a real-world, organisation-wide assessment of CSC within a large, global organisation providing services to organisations operating in safety-critical and critical-infrastructure settings.

**Practical implications**
Our study offers a practical, scalable way to assess CSC, generating meaningful insights despite industrial constraints. It enables organisations to benchmark maturity, identify gaps, and prioritise targeted improvements using workforce diversity as a guide.

## 1. Introduction

In today's interconnected, digital world, the threat of cyberattacks is not just a technical concern and instead has become an organisational challenge that can disrupt operations, endanger public safety, and inflict reputational and financial damage. The impacts of a cyberattack can extend beyond a single organisation and affect the wider ecosystem of customers, partners, and service providers it connects with. For example, the March 2019 attack on Norsk Hydro, a large aluminium and energy company, impacted approximately 35,000 employees across 40 countries, and resulted in $71 million in losses and damaged reputation (Briggs, 2019; Fotis, 2024). The breach occurred when an employee unknowingly opened a malicious email attachment from what appeared to be a trusted customer, demonstrating how a single human error can trigger major consequences (Chang *et al.*, 2025; Okolo *et al.*, 2021; Boyens *et al.*, 2022).

These cascading effects are not unique. Recent high-profile incidents have shown how cyberattacks can propagate across interconnected sectors and disrupt services essential to society. For example, the ransomware attack on the Colonial Pipeline, a major US fuel artery, in May 2021, caused significant disruptions to fuel distributions along the eastern seaboard (Wikipedia, 2021; Beerman *et al.*, 2023). The attackers gained access using a compromised password for an inactive Virtual Private Network account. The ransomware "DarkSide" targeted IT systems, leading to the shutdown of the pipeline to prevent further compromise. Colonial Pipeline paid a ransom of $4.4 million to regain access to their data and systems (Wikipedia, 2021; Beerman *et al.*, 2023). In February 2021, cyber actors gained unauthorised access to the Supervisory Control and Data Acquisition system of a US water treatment facility and attempted to increase the dosage of a hazardous chemical (*Compromise of U.S. Water Treatment Facility*, 2021), potentially leading to public health crisis. Although operators quickly detected and reversed the change, the cybersecurity (CS) incident exposed the risks, weak password practices, shared credentials, outdated operating systems, and insecure remote-access tools.

A common thread in these incidents is the pivotal role of human factors and technology alone is not sufficient to prevent these attacks. Verizon (2024) reports that 68% of data breaches were caused by human error, including social engineering scams (Abroshan *et al.*, 2021). A 2024 IBM (2025) report shows the average cost of a data breach reached a record high of $4.88 million. Human error and IT failures account for nearly half of all CS breaches, demonstrating that human factors contribute as much as vulnerable technological systems. A 2024 ProofPoint survey of 8550 users and security professionals across 15 countries found that 71% admitted to taking risky actions, 96% of whom were aware of the risk (*2024 State of the Phish*, 2024). Microsoft Digital Defense Report covering trends from July 2023 through June 2024 shows that their customers are facing 600 million attacks daily (*Microsoft Digital Defense Report: 600 million cyberattacks per day around the globe*).

As cyber threats increasingly exploit human behaviour (Amoresano and Yankson, 2023), organisations must foster resilient cybersecurity culture (CSC) to manage human-related vulnerabilities (Aksoy, 2024; Kannelønning and Katsikas, 2023). Reegård and Blackett (2019) define CSC as "the cybersecurity culture of the organisation that will influence the deployment and effectiveness of the cybersecurity management resources, policies, practices and procedures as they represent the work environment and underlying perceptions, attitudes, and habitual practices of employees at all levels" (p. 4041).

### *1.1. Creation and layers of organisational culture*

CSC is rooted in broader organisational culture, making it necessary to understand how culture is formed and transmitted within organisations in order to intentionally reshape it to improve CS practices (Listyowardojo *et al.*, 2014). Organisational culture develops through ongoing social interaction, as individuals engage with one another in everyday work contexts (Antonsen, 2009; Berger and Luckmann, 1991; Schein, 2010). Berger and Luckmann (1991) describe this process as one in which individuals externalise their interpretations of reality through verbal and non-verbal actions. Over time, these shared interpretations become stabilised as common beliefs and expectations, giving rise to recurring behavioural patterns and routines (i.e., institutionalisation). As these patterns are reinforced, they are internalised by individuals and come to be experienced as normal and self-evident ways of acting.

In the context of CS, such internalised routines may persist even when they increase vulnerability, particularly when individuals continue familiar practices without reflecting on alternative or more secure behaviours. When repeated over time, these practices can become taken for granted and resistant to change (Berger and Luckmann, 1991). For example, a new employee who observes colleagues sharing passwords may come to view this behaviour as acceptable and gradually adopt it unless corrective signals are introduced.

Schein's model of organisational culture conceptualises culture as comprising three interrelated layers: artefacts, espoused beliefs and values, and underlying assumptions (Schein, 2006; Schein, 2010). Artefacts represent visible behaviours and practices, such as locking a computer when leaving a workstation. Espoused beliefs and values reflect what individuals explicitly consider appropriate or inappropriate conduct, for example the belief that password sharing is unacceptable. Underlying assumptions operate at a deeper, largely unconscious level, shaping perceptions and behaviour through taken-for-granted understandings.

Returning to the example, the act of password sharing constitutes the observable artefact. Interpreting this behaviour as necessary to get work done reflects an espoused belief. The underlying assumption becomes apparent when the employee later receives an email appearing to come from a colleague requesting a password. Because password sharing has been normalised, the employee may be unlikely to question the request or consider the possibility of phishing. This illustrates how organisational culture is created, reproduced, and embedded in everyday practices that directly influence cybersecurity behaviour.

### 1.2. The gap and research questions

Despite growing academic interest in CSC, a critical gap remains in understanding how it is assessed and operationalised in real-world settings, particularly in organisations that work with or support safety-critical and critical-infrastructure sectors (Uchendu *et al.*, 2021; Chowdhury *et al.*, 2022). In these environments, cyberattacks can have implications that reach beyond financial loss, as disruptions affecting service providers may influence the operational reliability and safety of the systems and organisations they support (Chang *et al.*, 2025; Okolo *et al.*, 2021).

Our paper addresses that gap in CSC research by presenting CSC assessment in a large, global organisation supporting safety-critical and critical-infrastructure industries. To our knowledge, prior studies in this field have typically examined single departments, regional contexts, or focused on individual-level training evaluations. Our research draws on a global survey administered to all internal and external employees across the organisation's global operations, encompassing a diverse, multinational workforce engaged in both office-based and field-based roles. We examine how CSC varies across demographic factors such as employment type and contract, tenure, managerial role, age, and gender to identify champions and target groups for improvement. This approach highlights patterns of strength and vulnerability, offering data-driven insights to help organisational leaders prioritise and tailor improvement measures.

For scientific communities, our study contributes to moving beyond narrow case studies to connecting theory with a real-world study in sectors where CS failures can have significant operational or organisational consequences. Our findings offer analytical insights into where and why CSC differences arise, highlighting key demographic factors that shape CSC and informing future research on CS resilience in critical settings.

For industry practitioners, our study provides a practical and scalable approach to large-scale CSC assessment. Our approach focuses on collecting meaningful data while addressing common industrial constraints, such as limited resource availability and a KPI-driven workforce (Uchendu *et al.*, 2021). Organisations can also use this approach to benchmark CSC maturity, identify gaps, and prioritise targeted improvement measures. By examining workforce diversity, we also offer actionable guidance that leaders can use to strengthen CS resilience across their organisations.

## 2. Methodology

A quantitative, survey-based approach was employed to examine demographic patterns in CSC across a large and diverse global workforce. This approach enables systematic statistical comparison between groups and is well suited to studies conducted at organisational scale.

### 2.1. *The context*

The organisation in our study provides professional services to organisations that operate in safety-critical and critical-infrastructure settings. CSC remains essential because employee behaviour can influence the security and reliability of services delivered to its customers (Boyens *et al.*, 2022). In recent years, the organisation has placed increasing strategic emphasis on strengthening its CS posture reflected in, for example, its decision to commission a global assessment of its CSC (this study). The global survey was the first time being conducted in the organisation, and initiated as part of a broader, organisation-wide effort to assess and strengthen CSC, with participation being endorsed through leadership channels across organisational units.

### 2.2. *Ethical and data protection considerations*

Our study was conducted in accordance with the guidelines provided by Sikt, the Norwegian Agency for Shared Services in Education and Research, a public administrative body under the Ministry of Education and Research in Norway. Based on Sikt's criteria, our research did not require formal ethical review for approval or notification, as it exclusively involved the collection and analysis of anonymous data and no personal data was processed (Sikt - Norwegian Agency for Shared Services in Education and Research). At no point during the study was it possible to identify individual participants.

The data collection, analysis, and reporting methods ensured anonymity, confidentiality, and voluntary participation. The survey was anonymous, and we only used aggregated data in analysis and reporting. Participation in the survey was voluntary. The data was stored securely on restricted systems in compliance with organisational data protection guidelines. Participants were informed that their responses would not be able to be linked to them personally or to their units in evaluation.

The scope of our study focuses on demographic variations in CSC. Although the survey captured additional organisational data, variables such as country and organisational division were omitted to comply with the organisation's data-protection requirements and protect proprietary organisational structures. Focusing solely on demographic factors ensures both generalisability and the anonymity of employees.

### 2.3. *Survey*

#### 2.3.1. *Development of the questionnaire*

In developing the questionnaire, we utilised two validated questionnaires: The Human Aspects of Information Security Questionnaire (HAIS-Q) (Parsons *et al.*, 2017), and the Cybersecurity Scale (CS-S) (Arpaci and Sevinc, 2022). Additionally, we developed the "Governance" dimension to measure familiarity with the organisation's CS related policies, procedures, and guidelines, and the "Leadership" dimension to assess the extent to which CS is prioritised and promoted within the organisation, as these topics were not covered in HAIS-Q or CS-S. The HAIS-Q instrument was used to assess individual-level CS knowledge, attitudes, and self-reported behaviours across domains. The CS-S was developed to assess individuals' CS related practices and perceptions across six dimensions. For CS-S, we excluded all but the Integrity dimension to reduce survey length and avoid conceptual overlap with the HAIS-Q captures key aspects of individual CS behaviour and attitudes. The Integrity dimension addresses elements not explicitly emphasised in the HAIS-Q, namely individuals' adherence to principles of honesty, consistency, and trustworthiness in CS practices. We adapted the questionnaire items for clarity and contextual relevance by replacing the term "cyberspace" with "digital platforms," a term more commonly understood and aligned with the language used within the organisation. The questionnaire responses were scored on a 5-point Likert scale where 1 stood for "Strongly Disagree", 2 for "Disagree", 3 for "Neutral", 4 for "Agree", and 5 for "Strongly Agree". Several survey items were reversed (Supplementary_file_Appendix_A).

Prior to scaling globally, we conducted a small-scale pilot study to test the methodology, refine the instrument, and assess practical implementation (Van Teijlingen and Hundley, 2001). After the learning from the pilot, we scaled up the assessment, inviting all internal and external employees (N=21148), including those who had participated in the pilot. The pilot did not constitute a separate analytical sample. External employees were not directly employed by the organisation but worked on projects through third-party contracts, consultancies, or outsourcing arrangements.

Following this phase, the questionnaire underwent a refinement process to remove items judged to be redundant and less relevant and to refine the items. These modifications were grounded in a collaborative consultative process involving the project team, subject matter experts, and the management of the organisation being studied. This expert-led adaptation ensured that the final questionnaire remained theoretically grounded while maintaining high practical relevance to the organisation's unique needs, culture, and operational environment. Ultimately, the questionnaire was designed to minimize time away from chargeable and prioritised work, encouraging higher global response rates. The final questionnaire included nine dimensions and eight demographic items (Table I and Supplementary_file_Appendix_A). Internal consistency was assessed using Cronbach's alpha ($\alpha$). Because the global survey had strict limits on length to avoid disrupting chargeable and prioritised work, several dimensions were measured with only a few broad items, resulting in lower internal consistency coefficients ($\alpha$ = 0.23–0.67; Table I). These values therefore reflect a pragmatic tradeoff between psychometric depth and the operational reality of conducting an enterprise-wide global survey at scale. The results are reported for transparency, with the understanding that conciseness was essential for organisational viability, and some compromise in reliability was considered unavoidable.

Table I. Nine dimensions, their short description and Cronbach's α (N=35)

| Questionnaire | Dimensions | N | Short description | Cronbach's α |
|---|---|---|---|---|
| HAIS-Q | Password Management | 3 | Secure and responsible password practices | 0.431 |
| HAIS-Q | Email Use | 6 | Awareness and cautious behavior regarding the safety of email links | 0.408 |
| HAIS-Q | Internet Use | 3 | Careful evaluation and cautious use of websites and downloads for work | 0.484 |
| HAIS-Q | Social Media Use | 3 | Cautious and privacy-aware use of social media | 0.228 |
| HAIS-Q | Mobile Devices | 3 | Cautious handling of devices and sensitive information in public or unsecured environments | 0.321 |
| HAIS-Q | Incident Reporting | 4 | Proactive reporting of incidents and rule violations | 0.585 |
| CS-S | Integrity | 4 | Awareness of the risks and limitations of storing information on digital platforms | 0.522 |
| Own | Governance | 6 | Knowledge of CS policies and guidelines, and practice of document classification and access control | 0.671 |
| Own | Leadership | 3 | Leadership support and promotion of cybersecurity awareness | 0.589 |

#### *2.3.2. Survey participants and data collection*

All employees were invited to participate (N=21,148), comprising 74.04% internal and 25.96% external employees. The survey was distributed for internal employees via an internal platform already familiar to employees, allowing access via the secure intranet rather than through email links. This approach aligned with the organisation's antiphishing measures, avoided mixed messages about link safety, and helped build participant confidence, supporting higher response rates. In contrast, external employees were invited via anonymized Microsoft Forms, as they lacked access to the internal system. To mitigate security concerns, the invitation included a verification protocol instructing recipients to contact the Project Manager (first author) via Microsoft Teams or email to confirm the link's authenticity. The survey remained open throughout July and August 2025. While external employees received email reminders, internal participation was driven through the leadership teams across major organisational divisions to foster local leadership ownership of the results.

### *2.4. Survey analysis*

The survey responses from the internal platform and Forms datasets were merged and cleaned to ensure consistency. Reversed survey items were recorded so that higher scores consistently reflected more positive outcomes (1=low, 5=high). For the Leadership dimension, the specific item designated as "only for line managers" was filtered during the data cleaning stage to remove responses from individuals who identified as non-managers, ensuring the integrity of the dimension scores. Medians and interquartile ranges (*IQR*) were computed for each dimension and to summarise response distributions across all dimensions, including demographic characteristics. Statistical tests were used to find significant differences in survey responses between and within demographic groups as well as testing for a non-response bias (Table I and Supplementary_file_Appendix_A). A significance level (*p*-value) of .05 was chosen. To control for Type I error during multiple comparisons, the Holm correction was applied to all post-hoc results. All subgroups had more than 10 participants; thus no elimination of any subgroups was necessary to ensure statistical reliability and anonymity.

To select appropriate tests for demographic group comparisons, normality tests were performed for all dimensions using a Shapiro–Wilk test (*scipy.stats.shapiro*). The Kolmogorov–Smirnov test was used as a supplementary test for very large subgroup sizes (i.e., more than 5000 responses) (*scipy.stats.kstest*). Due to the large sample, normality tests were done with Python utilizing the SciPy (v1.16.0) library. For 97% of the groups, the normality test results showed that the data was not normally distributed. As a result, nonparametric methods were employed for all primary analyses.

The Kruskal–Wallis *H* test was used to determine if significant differences existed between demographic groups. Post-hoc pairwise comparisons (i.e., Dunn's tests) were performed only when the Kruskal-Wallis test results indicated significant differences. To evaluate the practical significance of the findings, the magnitude of effect for significant pairwise differences was calculated using the Rank-Biserial Correlation ($r_{rb}$). Furthermore, we tested for non-response bias by comparing early (responded on 1-14 July), middle (responded on 15 July-14 August), and late (responded on 15-31 August) participants groups. These analyses were computed using the Jeffreys's Amazing Statistics Programme (JASP) (Han and Dawson, 2020).

## 3. Results

The survey received a 30.75% response rate ($N = 6502$). A Kruskal–Wallis $H$ test was used to assess non-response bias across early, middle, and late respondents. Significant differences were found between groups for Email Use and Internet Use. For Email Use, the analysis showed a significant difference between groups, $H(2) = 6.59$, $p = .037$, with very small effect size (Rank $\eta^2 = 0.000$, 95% CI [$7.06\times10^{-4}$, 0.004]). Dunn's post-hoc tests with Holm correction indicated that middle respondents scored significantly higher than late respondents, $z = 2.56$, $p_{\text{holm}} = .031$, and the effect size was small ($r_{r\beta} = 0.049$). No other significant differences were found.

For Internet Use, the Kruskal–Wallis test also identified a significant group difference, $H(2) = 7.90$, $p = .019$, with very small effect size (Rank $\eta^2 = 0.000$, 95% CI [$9.08\times10^{-4}$, 0.003]). Dunn's tests showed that late respondents scored significantly higher than early respondents, $z = -2.50$, $p_{\text{holm}} = .037$, with a small effect size ($r_{r\beta} = 0.044$), while the remaining pairwise comparisons were not significant. Although statistically significant, the magnitude of non-response bias in both dimensions was evaluated as small and unlikely to meaningfully impact the validity of the overall findings.

The participants were predominantly internal permanent (78.78%), full-time (82.51%), direct recruits (79.01%), not line managers (84.31%), male (66.15%), older than 40 years old (67.66%), had a tenure for 1-5 years (32.64%) and 11-20 years (26.30%), and office-based (78.78%) (Table II). Password Management showed the highest median score with a relatively narrow IQR, followed by Mobile Devices and Social Media (Table III). In contrast, Email Use and Governance showed the lowest median score.

Table II. Demography information of participants and the frequency of each category of being in Group 1 and 2

| Demography item | Category | N | % | Frequency of being in Group 1 (Scoring higher across dimensions) | Frequency of being in Group 2 (Scoring lower across dimensions) |
|---|---|---|---|---|---|
| Employment contract | Internal permanent | 5122 | 78.78 % | 10 | 2 |
| | External fixed-term | 713 | 10.97 % | 0 | 8 |
| | Internal time-limited | 337 | 5.18 % | 4 | 4 |
| | External short-term | 330 | 5.08 % | 4 | 3 |
| | **TOTAL** | **6502** | 100% | | |
| Employment type | Full-time | 5365 | 82.51 % | 12 | 1 |
| | Task-based or project-based | 700 | 10.77 % | 5 | 4 |
| | Part-time | 437 | 6.72 % | 0 | 12 |
| | **TOTAL** | **6502** | 100% | | |
| Recruitment paths | Direct recruits | 5137 | 79.01 % | 5 | 4 |
| | Via M&A | 693 | 10.66 % | 11 | 0 |
| | Through an external agency or consultancy | 541 | 8.32 % | 0 | 6 |
| | Do not know | 131 | 2.01 % | 0 | 6 |
| | **TOTAL** | **6502** | 100% | | |
| Line managerial role | Not line managers | 5482 | 84.31 % | 0 | 7 |
| | Line managers | 1020 | 15.69 % | 7 | 0 |
| | **TOTAL** | **6502** | 100% | | |
| Gender | Male | 4301 | 66.15 % | 4 | 1 |
| | Female | 2055 | 31.61 % | 1 | 4 |
| | Prefer not to say | 135 | 2.08 % | 2 | 2 |
| | Other | 11 | 0.17 % | 0 | 0 |
| | **TOTAL** | **6502** | 100% | | |
| Age group | Older than 50 years old | 2646 | 40.70 % | 11 | 0 |
| | 41-50 years old | 1753 | 26.96 % | 7 | 1 |
| | 31-40 years old | 1328 | 20.42 % | 7 | 2 |
| | 25-30 years old | 528 | 8.12 % | 1 | 13 |
| | Prefer not to say | 139 | 2.14 % | 5 | 1 |
| | Younger than 25 years old | 108 | 1.66 % | 0 | 14 |
| | **TOTAL** | **6502** | 100% | | |
| Tenure | 1-5 years | 2122 | 32.64 % | 5 | 2 |
| | 11-20 years | 1710 | 26.30 % | 0 | 4 |
| | 21-30 years | 948 | 14.58 % | 2 | 2 |
| | 6-10 years | 781 | 12.01 % | 0 | 3 |
| | Less than 1 year | 659 | 10.14 % | 5 | 2 |
| | Longer than 30 years | 282 | 4.34 % | 2 | 1 |
| | **TOTAL** | **6502** | 100.00 % | | |
| Work base | Office-based employees | 4361 | 78.78 % | 3 | 2 |
| | Field-based employees (e.g., surveyors, auditors, assessors) | 2141 | 10.97 % | 2 | 3 |
| | **TOTAL** | **6502** | 5.18 % | | |

The Kruskal–Wallis *H* test revealed significant differences across all dimensions (Table III). Integrity was the only dimension in which the Kruskal–Wallis *H* test showed significant differences across all demographic factors. The Kruskal-Wallis effect sizes ($\eta^2_{(rank)}$) ranged from 0.001 to 0.019 across dimensions, indicating very small to small effects. The highest effect sizes were observed for Integrity (i.e., work base and contract), Leadership (i.e., contract and employment), Password Management (i.e., contract), Incident Reporting (i.e., managerial role), and Email Use (i.e., age group).

Dunn's post hoc tests revealed that fulltime participants, internal permanent participants, participants older than 50 years, and those recruited via M&A (Merge and Acquisition) showed the highest frequency of being in Group 1 (i.e., the higherscoring group) compared with their respective comparison groups across multiple dimensions (Tables II and Supplementary_file_Appendix_B). Another notable pattern was that line managers appeared exclusively in Group 1 and did not appear in Group 2 for any dimension.

In contrast, part-time participants and participants younger than 31 years showed the highest frequency of being in Group 2 (i.e., the lower-scoring group) compared with their respective comparison groups across multiple dimensions (Tables II and Supplementary_file_Appendix_B). Another pattern revealed that external fixed-term participants, those recruited through an external agency or consultancy, those who responded 'do not know' to their recruitment path, non-line managers, and participants with 6–20 years of tenure appeared exclusively in Group 2 and did not appear in Group 1 for any dimension.

In Dunn's posthoc tests, rankbiserial correlation ($r_{rb}$) values for effect sizes ranged from 0.033 to 0.299, suggesting very small to moderate effects (Supplementary_file_Appendix_B). The six highest effect sizes ($r_{rb} =$ 0.250 to 0.299) were observed for Email Use (i.e., age group), Integrity (i.e., gender), and Mobile Devices (i.e., age group). Specifically, in Email Use, participants older than 50 years had significantly higher scores than did those younger than 25 years, $z = 5.33$, $p < .001$, $p_{holm} < .001$, with a moderate effect size ($r_{r\beta} = .299$). Similarly, those who selected "Prefer not to say" had significantly higher scores than did the youngerthan25 group ($z = 3.92$, $p < .001$, $p_{holm} < .001$, $r_{r\beta} = .279$). Furthermore, the 41–50 years group had significantly higher scores than did the youngerthan25 group ($z = 4.43$, $p < .001$, $p_{holm} < .001$, $r_{r\beta} = .253$). In Integrity, participants in the "Prefer not to say" category in gender had significantly higher scores than did female participants ($z = 5.19$, $p < .001$, $p_{holm} < .001$, $r_{r\beta} = .267$). In Mobile Devices, both the 41–50 years group ($z = 4.76$, $p < .001$, $p_{holm} < .001$, $r_{r\beta} = .266$) and the older than50 group ($z = 4.56$, $p < .001$, $p_{holm} < .001$, $r_{r\beta} = .251$) had significantly higher scores than did those younger than 25 years.

Table III. Dimensions' Median (IQR) and summary of Kruskal-Wallis $H$ test

| **Dimension** | ***Median* (IQR)** | **Contract (df=3) (Rank η2)** | **Employment (df=2) (Rank η2)** | **Recruitment (df=3) (Rank η2)** | **Managerial role (df=1) (Rank η2)** | **Gender (df=1) (Rank η2)** | **Age group (df=5) (Rank η2)** | **Tenure (df=5) (Rank η2)** | **Work-base (df=1) (Rank η2)** |
|---|---|---|---|---|---|---|---|---|---|
| **Email Use** | **3.83 (0.67)** | $H$=11.439* (0.001) | n.s. | $H$=13.806** (0.002) | n.s. | n.s. | $H$=87.190*** (0.013) | $H$=23.952*** (0.003) | $H$=10.219** (0.001) |
| **Governance** | **3.83 (0.84)** | $H$=30.595*** (0.004) | $H$=21.828*** (0.003) | $H$=8.182* (7.974×10-4) | n.s. | $H$=23.975*** (0.003) | n.s. | $H$=29.256*** (0.004) | $H$=38.468*** (0.006) |
| **Incident Reporting** | **4.25 (0.50)** | n.s. | $H$=11.633** (0.001) | $H$=33.414*** (0.005) | $H$=93.414*** (0.014) | $H$=26.907*** (0.004) | $H$=50.300*** (0.007) | n.s. | n.s. |
| **Integrity** | **4.00 (0.75)** | $H$=111.985*** (0.017) | $H$=47.083* (0.007) | $H$=54.887*** (0.008) | $H$=35.223*** (0.005) | $H$=38.708*** (0.005) | $H$=12.532* (0.001) | $H$=12.702* (0.001) | $H$=123.048*** (0.019) |
| **Internet Use** | **4.00 (0.66)** | $H$=8.336* ($8.212\times10^{-4}$) | $H$=27.119*** (0.004) | $H$=12.796** (0.002) | $H$=5.184* (6.436×10-4) | n.s. | n.s. | $H$=30.723*** (0.004) | n.s. |
| **Leadership** | **4.00 (0.83)** | $H$=85.554*** (0.013) | $H$=104.651*** (0.016) | $H$=19.772* (0.003) | n.s. | $H$=15.598* (0.002) | $H$=14.391* (0.001) | $H$=14.908* (0.002) | $H$=4.977* (6.118×10-4) |
| **Mobile Devices** | **4.33 (1.00)** | $H$=18.126*** (0.002) | $H$=24.577*** (0.003) | $H$=12.849** (0.002) | $H$=35.716*** (0.005) | $H$=10.063* (0.001) | $H$=30.252*** (0.004) | n.s. | n.s. |
| **Password Management** | **5.00 (0.33)** | $H$=93.213*** (0.014) | $H$=23.778*** (0.003) | $H$=29.684*** (0.004) | $H$=18.266*** (0.003) | n.s. | $H$=37.607*** (0.005) | $H$=12.093* (0.001) | $H$=32.549*** (0.005) |
| **Social Media Use** | **4.33 (0.67)** | $H$=21.095*** (0.003) | $H$=25.039*** (0.004) | n.s. | $H$=8.209** (0.001) | n.s. | $H$=23.952*** (0.003) | $H$=12.929* (0.001) | n.s. |

*Note*. Numbers represent the Kruskal-Wallis $H$ statistic $*p < .05$ $**p < .01$ $***p < .001$. n.s.: not significant.

## 4. Discussions

Our findings indicate a broadly consistent CSC across the organisation, with modest demographic variations. Across demographic comparisons, six effects reached the moderate range. These findings suggest that employees younger than 25 years old may need additional support for the cautious handling of devices and sensitive information in public or unsecured environments (i.e., Mobile Devices) as well as for secure email practices (i.e., Email Use), as also found here (Branley-Bell *et al.*, 2022). For those who selected "prefer not to say" for gender category scored significantly higher than female employees in Integrity. They may demonstrate greater privacy sensitivity, an attitude that aligns with the Integrity dimension's focus on recognising  the risks and limitations of storing information on digital platforms.

Another key finding is that although statistically significant differences were identified across all demographic variables, effect sizes were very small to moderate, suggesting subtle rather than substantial group differences. Nevertheless, in a large organisation supporting safety-critical and critical infrastructure industries, even small behavioural variations may have meaningful operational implications (Chowdhury *et al.*, 2022; Sikha and Somepalli, 2023; Boyens *et al.*, 2022). CSC improvement efforts can therefore be targeted toward specific groups while maintaining organisation-wide initiatives, as tailoring CS awareness has been shown to strengthen CS posture (Delso-Vicente *et al.*, 2025).

A consistent scoring pattern emerged in which full-time, internal permanent, older employees (i.e., 50+ years old), and M&A recruits demonstrated higher CSC scores across multiple dimensions (Table II and Supplementary_file_Appendix_B). These findings align with prior research showing that CS behaviours strengthen with age (Branley-Bell *et al.*, 2022), with the 60-69 age group showing the strongest CS behaviours (de Bruin and Mersinas, 2024), and permanent, full time employment (Sharma and Warkentin, 2019; Anwar, He and Yuan, 2016). Full-time, internal permanent employees generally have more consistent exposure to organisational policies, training, and communication, fostering stronger awareness of CS practices compared to part-time or external employees. Older employees may also have more professional experience, along with greater exposure to CS training or incidents (Handri, Sensuse and Lusa, 2024), contributing to higher awareness and more cautious behavior. M&A recruits may bring established CS practices from previous organisations, especially if recruited from organisations providing CS improvement services,, and may also receive targeted onboarding and integration, enhancing CS awareness. Recent acquisitions of organisations focused on improving CS by the organisation in our study may have further influenced this finding.

Our findings show that line managers consistently scored higher than non-managers across seven of nine dimensions, indicating a relatively stronger understanding of CS expectations. Our findings are diverse from findings in Kannelønning & Katsikas's (2023) study that reports cases where managers show lower and inconsistent CS awareness. These findings may be because, in the organisation being studied, line managers are typically responsible for their team members' performance including their CS behaviours. This may also reflect increased prioritisation of CS improvement by the organisation's top management, which is suggested as the strongest influence to CSC (Huang and Pearlson, 2019; Delso-Vicente *et al.*, 2025).

These findings suggest that job stability and experience (Sharma and Warkentin, 2019; Baltuttis, Teubner and Adam, 2024), and leadership responsibilities support a more desirable CSC (Huang and Pearlson, 2019). These groups may therefore function as potential culture carriers or champions in efforts to help build stronger CSC by, for example, explicitly role modeling CS behaviours and practices, initiating dialogues, and actively participating in CS initiatives and hubs (Alshaikh, 2020).

In contrast, part-time employees and participants younger than 31 years old show a consistent pattern of lower scoring across dimensions. Our findings reveal that external fixed-term participants, those recruited through an external agency or consultancy, those who responded 'do not know' to their recruitment path, non-line managers, and participants with 6–20 years of tenure consistently scored lower than their respective comparison groups (Table II and Supplementary_file_Appendix_B), and never higher, indicating a relatively lower understanding of CS practices. As previously mentioned, these findings are consistent with previous research that found part-time, external, and younger employees to be less committed to desirable CS practices (Anwar, He and Yuan, 2016; Sharma and Warkentin, 2019; de Bruin and Mersinas, 2024). Furthermore, part-time and external employees are usually expected to prioritise chargeable work, leaving them with limited time or organisational support to participate in CS-related training or awareness activities. Younger participants may have less professional experience, fewer opportunities to participate in CS-related training, or may not yet have developed the cautious behaviors that come with greater tenure and exposure to CS risks (Baltuttis, Teubner and Adam, 2024). Employees with 6–20 years of tenure may rely on established habits, and without regular refreshers or role-based reinforcement, these routines can become outdated and misaligned with current CS expectations.

These lower-scoring groups may be more susceptible to common CS threats and therefore require tailored measures (Delso-Vicente *et al.*, 2025), such as enhanced onboarding, targeted training, mentorship from higher-scoring groups, or clearer communication of expectations (European Union Agency for Network and Information Security, 2017). The mid-tenure group, for example, may benefit from targeted initiatives such as refreshed training content that better aligns with how mid-tenure employees work, more frequent reinforcement of key messages, and clearer articulation of expectations around CS responsibilities for individual contributors (European Union Agency for Network and Information Security, 2017). External and part-time employees may strengthen their CSC through more targeted onboarding support, increased access to essential CS policies and guidelines, more frequent and role-relevant awareness activities, and clearer communication channels (Alshaikh, 2020). Embedding CS expectations in working agreements and leveraging high-scoring groups as champions can further reinforce desired CS practices (Sharma and Warkentin, 2019).

The findings also show patterns in which certain groups scoring higher in some dimensions and lower in others, such as employees with task- or project-based employment, direct recruits, external short-term and internal time-limited. These mixed scoring groups highlight the need for consistency reinforcement. Improvement measures for this mixed-scoring group should be nuanced and targeted, addressing specific gaps while leveraging existing strengths within these groups (Delso-Vicente *et al.*, 2025). Task- or project-based, internal time-limited, and external short-term employees may have irregular schedules and/or variable exposure to organisational policies and training, which can result in stronger awareness in areas closely related to their project work, but less familiarity with broader or less frequently encountered CS practices. Furthermore, employees from acquired organisations with strong CS expertise often bring more advanced CS knowledge and mature practices, whereas direct recruits may lack this deeper CS experience. This suggests that preserving and leveraging acquired CS-focused organisations is more advantageous rather than subsuming them entirely into the host organisation's practices (Schraeder and Self, 2003). Facilitating two-way learning between acquired and acquiring organisations can enrich the overall CSC and promote best practices across the organisation (Dhillon, Syed and Pedron, 2016).

### *4.1. Practical Implications*

Based on the findings, we have identified key improvement suggestions to improve the CSC of the organisation. First, strengthen and leverage existing organisational CS measures and initiatives. The broadly consistent CSC observed across the organisation indicates that these measures have been effective in establishing a common baseline. Prioritising targeted adjustments based on demographic patterns in existing CS training and communication are likely to lead to further meaningful improvements.

Second, demographic patterns identified signal the need for targeted, tailored improvement measures. Although CSC is observed as relatively consistent, little variations can have operational consequences. Improvement measures and initiatives thus need to be tailored and targeted to the target group, while leveraging culture carrier groups.

Third, formalise line managers as CSC carriers. This can be done by, for example, strengthening expectations for line managers to reinforce CS behaviours and practices in daily work, and integrating CSC responsibilities into existing routines such as CS moments in team meetings.

Fourth, strengthen and standardise the onboarding process to emphasise CS expectations. CS onboarding requirements, which may include CS training, need to be extended to all employment types, including part-time, external, and time-limited employees, to ensure a more consistent onboarding outcome across the workforce. To reduce knowledge decay, these expectations should also be reinforced beyond onboarding through recurring, role-relevant reminders and touchpoints (e.g., periodic refreshers, brief CS moments in team meetings, and consistent messaging in operational communications).

Five, adapt CSC improvement measures to accommodate irregular schedules, limited access to internal systems, and chargeable work. For example, by embedding CS expectations more explicitly into working agreements, project documentation, and contractual arrangements.

Sixth, leverage CS expertise from M&A recruits. CS expertise needs to be preserved and leveraged from acquired organisations, enabling two-way learning mechanisms to foster more mature CS practices across the organisation.

Seventh, target younger and early-career employees, especially on secure handling of devices and sensitive information in public places and secure email practices.

### *4.2. Limitations*

The relatively low response rate may introduce bias, but non-response analyses indicated negligible differences, suggesting the sample remained broadly representative. Data-protection constraints prevented inclusion of contextual variables (e.g., organisational divisions and countries) and restricted triangulation. We thus focused on demographic factors that could be analysed meaningfully. Self-reported data may introduce bias and some dimensions showed lower internal consistency, which we address through transparent reporting and recommend strengthening through future longitudinal work, deeper scales, and triangulation when permissible. Although some subgroups were small, all met the minimum threshold for statistical reliability, helping preserve the validity of subgroup comparisons.

### *4.3. Future Directions*

Future research should explore CSC longitudinally, particularly as targeted improvement measures are implemented. Incorporating additional organisational variables such as organisational divisions, geographical region, or team level characteristics would also provide deeper contextual insight and help identify structural drivers of CSC differences. Methodologically, future work could strengthen psychometric robustness through refined scales and factor validation while maintaining survey conciseness. Integrating objective behavioural data (e.g., phishing simulations or incident reporting) and mixed-methods approaches may further clarify the mechanisms underlying demographic CSC differences and inform more effective interventions.

## 5. Conclusions

Our findings contribute both to the organisation's improvement efforts and to the broader understanding of how CSC manifests in large, global organisations. Although effect sizes were small, consistent patterns emerged across job stability, managerial responsibility, age, recruitment background, and tenure. These variations are subtle rather than structural, indicating a broadly aligned CSC that provides a stable foundation for targeted improvement.

By presenting results by demographic group, the study offers actionable insights for leaders: lower-scoring groups can be prioritised for tailored interventions, while higher-scoring groups can act as role models and culture carriers. Recognising workforce diversity as a driver of CSC can support clearer CS communication, more effective training, and stronger leadership engagement.

# Appendices

Appendix A. The questionnaire items

| Type of item | Category | Survey items |
|---|---|---|
| Demographic items | My employment contract is: | Internal permanent (Directly employed by the organisation with no end date) |
| | | Internal time-limited (Directly employed by the organisation for a set period) |
| | | External fixed-term (Hired via external provider for a set period) |
| | | External short-term (Externally hired for brief, temporary tasks) |
| | My employment is: | Full-time |
| | | Part-time |
| | | Task-based or project-based (irregular schedule) |
| | I was recruited: | Directly by the organisation |
| | | Via M&A |
| | | Through an external agency or consultancy |
| | | Do not know |
| | My role is: | I am a line manager |
| | | I am NOT a line manager |
| | Gender | Female |
| | | Male |
| | Age group | Younger than 25 years old |
| | | 25-30 years old |
| | | 31-40 years old |
| | | 41-50 years old |
| | | Older than 50 years old |
| | | Prefer not to say |
| | I have been working for the organisation for: | Less than 1 year |
| | | 1-5 years |
| | | 6-10 years |
| | | 11-20 years |
| | | 21-30 years |
| | | Longer than 30 years |
| | I am predominantly: | Field-based |
| | | Office-based |
| CSC survey items | Email Use (N=6) | Links in emails help me get things done faster. |
| | | It's always safe to click on links in emails from colleagues. |
| | | I don't always click on links in emails just because they come from colleagues. |
| | | Not using links in emails is necessary to keep the organisation secure. |
| | | It's always safe to click on links in emails from people I know. |
| | | Sometimes I'm unsure if an email link is safe to click. |
| | Governance (N=6) | I log out of systems and tools when I no longer need access to them. |
| | | I regularly check who has access to my digital files. |
| | | I feel confident using the organisation's cybersecurity policies to guide my actions in case of a security issue. |
| | | I know where to find the organisation's cybersecurity policies and guidelines. |
| | | It's impractical to classify documents into "unlabeled," "open," "restricted," "confidential," or "secret". |
| | | I always classify documents in my daily work (e.g. Open, confidential) |
| | Incident Reporting (N=4) | I believe that all security incidents should be reported, even if they seem minor. |
| | | If I see a colleague ignoring security rules, I take action or report it. |
| | | If I notice someone acting suspiciously at work, I would report it. |
| | | It's optional to report security incidents. |
| | Integrity (N=4) | I believe sharing information on digital platforms is always safe. |
| | | It's possible for third parties to access information stored on digital platforms. |
| | | Storing data on digital platforms is not always secure. |
| | | Information I have stored on digital platforms may be lost or deleted. |
| | Internet Use (N=3) | I assess the safety of websites before entering information. |
| | | When accessing the internet at work, I visit any website that I want. |
| | | I download any files onto my work computer that will help me get the job done. |
| | Leadership (N=3) | My line manager models and promotes cybersecure practices |
| | | My organisational division has prioritised cybersecurity awareness in the past 12 months. |
| | | ONLY FOR MANAGERS: As a manager, I feel I have the tools and resources needed to promote cybersecure practices in my team. |
| | Mobile Devices (N=3) | When working in public places, I always keep my laptop with me. |
| | | It's risky to send sensitive work files using a public Wi-Fi network. |
| | | I check that strangers can't see my laptop screen if I'm working on a sensitive document. |
| | Password Management (N=3) | It's a bad idea to share my work passwords, even if a colleague asks for it. |
| | | It's safe to use the same password for social media and work accounts. |
| | | I am allowed to share my work passwords with colleagues. |
| | Social Media Use (N=3) | I avoid posting things on social media that I would not say in a public setting. |
| | | I feel free to post whatever I want about work on social media. |
| | | I regularly review the privacy settings on my social media accounts (e.g. Linkedin). |

Appendix B. Dunn's post-hoc pairwise comparisons across demographic groups

| Dimension | Variable | Group 1 (Scoring higher) | Group 2 (Scoring lower) | Z | $r_{rb}$ | *p*-value | $p_{holm}$ |
|---|---|---|---|---|---|---|---|
| Email Use | Employment contract | Internal permanent | External fixed-term | 3.367 | 0.078 | < .001 | 0.005 |
| | Age group | 31-40 years old | 25-30 years old | 3.748 | 0.113 | < .001 | 0.002 |
| | | 41-50 years old | 25-30 years old | 5.498 | 0.157 | < .001 | < .001 |
| | | Older than 50 years old | 25-30 years old | 7.488 | 0.204 | < .001 | < .001 |
| | | Prefer not to say | 25-30 years old | 3.529 | 0.187 | < .001 | 0.003 |
| | | 31-40 years old | Younger than 25 years old | 3.587 | 0.209 | < .001 | 0.003 |
| | | Older than 50 years old | 31-40 years old | 4.878 | 0.095 | < .001 | < .001 |
| | | 41-50 years old | Younger than 25 years old | 4.428 | 0.253 | < .001 | < .001 |
| | | Older than 50 years old | Younger than 25 years old | 5.327 | 0.299 | < .001 | < .001 |
| | | Prefer not to say | Younger than 25 years old | 3.918 | 0.279 | < .001 | < .001 |
| | | Older than 50 years old | 41-50 years old | 2.726 | 0.048 | 0.006 | 0.038 |
| | Tenure | 21-30 years | Less than 1 year | 3.287 | 0.095 | 0.001 | 0.014 |
| | | 21-30 years | 1-5 years | 3.104 | 0.070 | 0.002 | 0.024 |
| | | Longer than 30 years | 1-5 years | 3.108 | 0.113 | 0.002 | 0.024 |
| | Work base | Field-based | Office-based | 3.197 | 0.048 | 0.001 | 0.001 |
| Governance | Employment contract | Internal time-limited | External fixed-term | 2.947 | 0.160 | 0.003 | 0.013 |
| | | Internal time-limited | External fixed-term | 3.422 | 0.160 | < .001 | 0.003 |
| | | Internal time-limited | Internal permanent | 5.172 | 0.167 | < .001 | < .001 |
| | Employment type | Full-time | Part-time | 4.643 | 0.133 | < .001 | < .001 |
| | | Task-based or project-based | Part-time | 3.844 | 0.138 | < .001 | < .001 |
| | Gender | Male | Female | 4.557 | 0.070 | < .001 | < .001 |
| | Tenure | 1-5 years | 6-10 years | 3.164 | 0.075 | 0.002 | 0.020 |
| | | Less than 1 year | 11-20 years | 4.188 | 0.111 | < .001 | < .001 |
| | | Less than 1 year | 6-10 years | 4.596 | 0.142 | < .001 | < .001 |
| | | Less than 1 year | Longer than 30 years | 3.094 | 0.128 | 0.002 | 0.024 |
| | Work base | Field-based | Office-based | 6.202 | 0.094 | < .001 | < .001 |
| Incident Reporting | Employment type | Full-time | Part-time | 3.399 | 0.097 | < .001 | 0.002 |
| | | Task-based or project-based | Part-time | 2.753 | 0.092 | 0.006 | 0.012 |
| | Recruitment paths | Via M&A | Do not know | 3.208 | 0.176 | 0.001 | 0.005 |
| | | Direct recruits | Through an external agency or consultancy | 2.993 | 0.077 | 0.003 | 0.008 |
| | | Via M&A | Direct recruits | 4.349 | 0.101 | < .001 | < .001 |

| Dimension | Variable | Group 1 (Scoring higher) | Group 2 (Scoring lower) | Z | $r_{rb}$ | *p*-value | $p_{holm}$ |
|---|---|---|---|---|---|---|---|
| | | Via M&A | Through an external agency or consultancy | 5.426 | 0.176 | < .001 | < .001 |
| | Line managerial | Line managers | Not line managers | 9.665 | 0.188 | < .001 | < .001 |
| | Gender | Male | Female | 4.642 | 0.071 | < .001 | < .001 |
| | Age group | 31-40 years old | 25-30 years old | 2.950 | 0.087 | 0.003 | 0.038 |
| | | 41-50 years old | 25-30 years old | 4.631 | 0.133 | < .001 | < .001 |
| | | Older than 50 years old | 25-30 years old | 6.288 | 0.170 | < .001 | < .001 |
| | | Older than 50 years old | 31-40 years old | 4.399 | 0.084 | < .001 | < .001 |
| Integrity | Employment contract | Internal permanent | External fixed-term | 8.142 | 0.187 | < .001 | < .001 |
| | | Internal permanent | External short-term | 4.579 | 0.148 | < .001 | < .001 |
| | | Internal permanent | Internal time-limited | 6.414 | 0.207 | < .001 | < .001 |
| | Employment type | Full-time | Part-time | 3.788 | 0.109 | < .001 | < .001 |
| | | Full-time | Task-based or project-based | 6.050 | 0.139 | < .001 | < .001 |
| | Recruitment paths | Via M&A | Do not know | 3.518 | 0.194 | < .001 | 0.002 |
| | | Via M&A | Direct recruits | 6.866 | 0.159 | < .001 | < .001 |
| | | Via M&A | Through an external agency or consultancy | 6.327 | 0.211 | < .001 | < .001 |
| | Line managerial | Line managers | Not line managers | 5.935 | 0.116 | < .001 | < .001 |
| | Gender | Male | Female | 4.421 | 0.068 | < .001 | < .001 |
| | | Prefer not to say | Female | 5.186 | 0.267 | < .001 | < .001 |
| | | Prefer not to say | Male | 3.915 | 0.195 | < .001 | < .001 |
| | Age group | Prefer not to say | 25-30 years old | 3.278 | 0.181 | 0.001 | 0.016 |
| | Tenure | Longer than 30 years | Less than 1 year | 3.446 | 0.141 | < .001 | 0.009 |
| | Work base | Office-based | Field-based | 11.093 | 0.168 | < .001 | < .001 |
| Internet Use | Employment type | Full-time | Part-time | 4.229 | 0.120 | < .001 | < .001 |
| | | Full-time | Task-based or project-based | 3.420 | 0.078 | < .001 | < .001 |
| | Recruitment paths | Via M&A | Direct recruits | 3.486 | 0.080 | < .001 | 0.003 |
| | Line managerial | Line managers | Not line managers | 2.277 | 0.044 | 0.023 | 0.023 |
| | Tenure | 1-5 years | 11-20 years | 4.249 | 0.079 | < .001 | < .001 |
| | | 1-5 years | 21-30 years | 3.132 | 0.069 | 0.002 | 0.021 |
| | | Less than 1 year | 11-20 years | 4.041 | 0.107 | < .001 | < .001 |
| | | Less than 1 year | 21-30 years | 3.343 | 0.096 | < .001 | 0.011 |
| Leadership | Employment contract | Internal permanent | External fixed-term | 7.069 | 0.160 | < .001 | < .001 |
| | | Internal permanent | External short- term | 6.539 | 0.160 | < .001 | < .001 |
| | | Internal time-limited | External short-term | 3.842 | 0.170 | < .001 | < .001 |
| | Employment type | Full-time | Task-based or project-based | 8.972 | 0.204 | < .001 | < .001 |

| Dimension | Variable | Group 1 (Scoring higher) | Group 2 (Scoring lower) | Z | $r_{rb}$ | *p*-value | $p_{holm}$ |
|---|---|---|---|---|---|---|---|
| | | Full-time | Part-time | 5.729 | 0.161 | < .001 | < .001 |
| | Recruitment paths | Direct recruits | Do not know | 3.894 | 0.194 | < .001 | < .001 |
| | | Via M&A | Do not know | 3.600 | 0.198 | < .001 | 0.002 |
| | Gender | Female | Prefer not to say | 2.802 | 0.142 | 0.005 | 0.025 |
| | | Male | Prefer not to say | 3.280 | 0.161 | 0.001 | 0.006 |
| | Age group | 31-40 years old | Prefer not to say | 3.289 | 0.161 | 0.001 | 0.015 |
| | Tenure | 1-5 years | 6-10 years | 2.972 | 0.070 | 0.003 | 0.044 |
| | Work base | Office-based | Field-based | 2.231 | 0.033 | 0.026 | 0.026 |
| Mobile Devices | Employment contract | Internal permanent | External fixed-term | 3.760 | 0.085 | < .001 | 0.001 |
| | Employment type | Full-time | Part-time | 4.926 | 0.138 | < .001 | 0.001 |
| | | Task-based or project-based | Part-time | 3.353 | 0.114 | < .001 | 0.002 |
| | Recruitment paths | Direct recruits | Through an external agency or consultancy | 2.850 | 0.073 | 0.004 | 0.026 |
| | Line managerial | Line managers | Not line managers | 5.976 | 0.115 | < .001 | < .001 |
| | Age group | 41-50 years old | 25-30 years old | 3.057 | 0.086 | 0.002 | 0.025 |
| | | 25-30 years old | Younger than 25 years old | 3.032 | 0.186 | 0.002 | 0.025 |
| | | 31-40 years old | Younger than 25 years old | 4.283 | 0.244 | < .001 | < .001 |
| | | 41-50 years old | Younger than 25 years old | 4.760 | 0.266 | < .001 | < .001 |
| | | Older than 50 years old | Younger than 25 years old | 4.562 | 0.251 | < .001 | < .001 |
| | | Prefer not to say | Younger than 25 years old | 3.288 | 0.236 | 0.001 | 0.012 |
| Password Management | Employment contract | Internal permanent | External fixed-term | 7.056 | 0.143 | < .001 | < .001 |
| | | Internal permanent | External short-term | 2.476 | 0.069 | 0.013 | 0.040 |
| | | External short-term | Internal time-limited | 3.267 | 0.115 | 0.001 | 0.004 |
| | | Internal permanent | Internal time-limited | 6.998 | 0.201 | < .001 | < .001 |
| | Employment type | Full-time | Part-time | 3.238 | 0.081 | 0.001 | 0.002 |
| | | Full-time | Task-based or project-based | 3.932 | 0.080 | < .001 | < .001 |
| | Recruitment paths | Direct recruits | Do not know | 3.092 | 0.138 | 0.002 | 0.006 |
| | | Via M&A | Do not know | 3.845 | 0.186 | < .001 | < .001 |
| | | Direct recruits | Through an external agency or consultancy | 3.619 | 0.083 | < .001 | 0.001 |
| | | Via M&A | Direct recruits | 2.292 | 0.047 | 0.022 | 0.044 |
| | | Via M&A | Through an external agency or consultancy | 4.468 | 0.129 | < .001 | < .001 |
| | Line managerial | Line managers | Not line managers | 4.274 | 0.074 | < .001 | < .001 |
| | Age group | 31-40 years old | 25-30 years old | 3.134 | 0.081 | 0.002 | 0.016 |

| Dimension | Variable | Group 1 (Scoring higher) | Group 2 (Scoring lower) | Z | $r_{rb}$ | *p*-value | $p_{holm}$ |
|---|---|---|---|---|---|---|---|
| | | 41-50 years old | 25-30 years old | 3.711 | 0.094 | < .001 | 0.002 |
| | | Older than 50 years old | 25-30 years old | 4.341 | 0.105 | < .001 | < .001 |
| | | 31-40 years old | Younger than 25 years old | 3.994 | 0.202 | < .001 | < .001 |
| | | 41-50 years old | Younger than 25 years old | 4.262 | 0.215 | < .001 | < .001 |
| | | Older than 50 years old | Younger than 25 years old | 4.536 | 0.222 | < .001 | < .001 |
| | | Prefer not to say | Younger than 25 years old | 3.556 | 0.225 | < .001 | 0.004 |
| | Job predominant role | Office-based | Field-based | 5.705 | 0.076 | < .001 | < .001 |
| Social Media Use | Employment contract | External short-term | External fixed-term | 3.068 | 0.110 | 0.002 | 0.011 |
| | | External short-term | Internal permanent | 4.496 | 0.146 | < .001 | < .001 |
| | | External short-term | Internal time-limited | 2.833 | 0.117 | 0.005 | 0.018 |
| | Employment type | Full-time | Part-time | 1.999 | 0.057 | 0.046 | 0.046 |
| | | Task-based or project-based | Full-time | 4.381 | 0.100 | < .001 | < .001 |
| | | Task-based or project-based | Part-time | 4.519 | 0.155 | < .001 | < .001 |
| | Tenure | 1-5 years | 11-20 years | 3.027 | 0.055 | 0.002 | 0.037 |
| | Line managerial | Line managers | Not line managers | 2.865 | 0.055 | 0.004 | 0.004 |
| | Age group | Older than 50 years old | 25-30 years old | 3.655 | 0.100 | < .001 | 0.004 |
| | | Older than 50 years old | Younger than 25 years old | 3.169 | 0.176 | 0.002 | 0.021 |

*Note*. Only comparisons reaching statistical significance after Holm adjustment are displayed.